\documentclass[12pt,preprint]{aastex}

\def\gta{ \lower .75ex \hbox{$\sim$} \llap{\raise .27ex \hbox{$>$}} }
\def\lta{ \lower .75ex\hbox{$\sim$} \llap{\raise .27ex \hbox{$<$}} }
\def \Ol {$\Omega_\Lambda$}
\def \Om {$\Omega_{\rm M}$}
\def\chs{\mbox{$\chi^2$}}
\newcommand{\OmegaL}{\Omega_{\Lambda}}
\newcommand{\kmsmpc}{{\rm \, km\, s}^{-1}{\rm Mpc}^{-1}}

\begin{document}

\title{Gamma Ray Bursts: Complementarity to Other Cosmological Probes}

\author{Meng Su\footnote{sxsumeng@pku.edu.cn}, Hong
Li\altaffilmark{1,2}, Zuhui Fan\altaffilmark{1}, and Bo
Liu\altaffilmark{3}} \affil{1 Department of Astronomy, School of
Physics, Peking University, Beijing, 100871, P. R. China} \affil{2
Institute of High Energy Physics, Chinese Academy of Sciences, P.O.
Box 918-4, Beijing, 100049, P. R. China}\affil{3 National Laboratory
for Information Science and Technology, Tsinghua University,
Beijing, 100084, P. R. China}

\setcounter{footnote}{0}

\begin{abstract}

We combine recent long Gamma Ray Bursts (GRBs) sample including 52
objects out to z$=$6.3 compiled from Swift Gamma Ray Bursts by
Schaefer (2006) with Type Ia Supernova (SNIa), Cosmic Microwave
Background (CMB) and Baryon Oscillation (BAO) to constrain
cosmologies. We study the constraints arising from GRBs alone and
the complementarity of GRBs to other cosmological probes. To analyze
the cosmological role of GRBs, we adopt the Hubble constant as a
free parameter with a prior in the range of
$H_0=72\pm8\enspace\kmsmpc$ instead of a fixed $H_0$ used in
previous studies. By jointly using SNIa gold/SNLS samples and GRBs,
the constraints on \Om -\Ol~ parameter space are dramatically
improved in comparing with those from SNIa data alone. The
complementarity of GRBs is mostly from the sub-sample with z$>$1.5
due to the different parameter degeneracies involved in luminosity
distances at different redshifts. Including GRB data in addition to
SNIa, BAO and CMB in our analysis, we find that the concordance
model with \Om $\sim$0.3 and \Ol$\sim$0.7 is still well within
1$\sigma$ confidence range.

\end{abstract}


\section{Introduction}

As the three famous observational supports for the Big Bang
Cosmology, there are three cosmological "pillars" to support the
picture of the accelerating universe: Observations of Type Ia
Supernova (SNIa), Cosmic Microwave Background (CMB) and Large Scale
Structures of the universe (LSS). According to the combined analysis
with all the available observational data, a concordance
cosmological model has been obtained (Riess et al. 2004; Astier et
al. 2006; Spergel et al. 2006; Tegmark et al. 2006). In the context
of Einstein's general relativity, a special sort of building blocks
of our universe, named \emph{dark energy}, needs to be introduced
(Weinberg 1989; Peebles \&Ratra 2003; Copeland et al.
2006)\footnote{Another possibility is so-called Modified Gravity.
See e.g. Lue (2006) for a recent review for DGP model and Song et
al. (2006) for the structure formation within this scenario.}. It
began to dominate our universe recently and probably hold the key
for the future evolution of the universe. Understanding the nature
of dark energy has become one of the most alluring and challenging
tasks with profound implications for both cosmology and theoretical
physics.

Due to the fundamental importance of dark energy, it is with great
motivation to use different cosmological probes to study the
property of dark energy with their own advantages. Comparing
constraints from different observations can provide important
consistency checks of cosmological models, and find potential
systematics of each probe, or possibly get inspiration of new
physics behind the bias. On the other hand, because of the different
parameter degeneracies involved in different observables, combining
proper means can greatly break sorts of degeneracies and shrink the
allowed parameter space so that to achieve the goal of precision
cosmology.

In particular, it has been known for a long time that standard
candles spanning a wide range of redshift can be used to tightly
constrain cosmological parameters. Since they carry integrated
information between sources and observers, we need standard candles
from diverse redshifts to understand tomographically the history of
the cosmic expansion and further to distinguish different
cosmological models. SNIa has been the most successful distance
indicator for cosmological studies. However, it is difficult to
observe very high redshift SNIa. On the other hand, long gamma-ray
bursts (GRBs) have been known as the most powerful events happened
in the universe. After the discovery of the X-ray afterglow of
GRB970228 (Costa et al. 1997), the major breakthrough in
understanding GRBs, a great effort has been dedicated to find tight
relations between observables that could potentially "calibrate"
GRBs as independent probes in cosmology (Amati et al. 2002;
Ghirlanda et al. 2004a; Liang \& Zhang 2005; Firmani et al. 2005;
Lamb et al. 2005; Wang \& Dai 2006; Firmani et al. 2006a; Firmani et
al. 2006b). There are several conspicuous advantages of GRB to be a
unique means for cosmological studies. The huge emitted power of
GRBs makes them detectable at z$\sim$20 or even higher, deep within
the range of the epoch of reionization (Schaefer 2003; Ghirlanda et
al. 2004b; Dai, et al. 2004, Hooper \& Dodelson 2006). They are dust
absorption-free tracers of the massive star formation in the
universe (Pacsynski 1998; Blain \& Natarajan 2000, Lamb \& Reichart
2000; Bromm \& Loeb 2002; Lloyd-Ronning et al. 2002; Firmani et al.
2004; Friedman \& Bloom 2005; Schaefer 2006). However, because of
the large dispersion of energies (Bloom et al. 2003) and the
difficulties to determine their redshifts (Pelangeon et al. 2006 and
references therein), GRBs have not been considered as veracious
lighthouse for a long time.


Recently, Ghirlanda et al.(2004a) found that for GRBs the total
energy emitted in $\gamma$--rays ($E_\gamma$), after a proper
collimated correction, correlates tightly with the peak spectral
energy $E_{\rm peak}$ (in a $\nu F_\nu$ plot). Thus, the
isotropically equivalent burst energy could be determined accurately
enough to be used practically for cosmological studies. Firmani et
al. (2006a) discovered a new correlation among emission properties
of GRBs, which involves the isotropic peak luminosity $L_{\rm iso}$,
the peak energy $E_{\rm peak}$ in the spectrum, and the "high
signal" time-scale $T_{0.45}$ used to characterize the variability
of GRBs. In the source rest-frame, this relation follows as: $L_{\rm
iso} = \tilde{K} E_{\rm peak}^{1.62} T_{0.45}^{ -0.49} $, with
$\tilde{K}$ being a constant. GRBs are thus becoming feasible
distance indicators. As they are the unique probe of high redshift
universe, GRBs might play important roles in dark energy studies,
especially if dark energy exhibits interesting behaviors at high
redshifts (but also see Linder \& Huterer 2003; Aldering et al.
2006).

In this \emph{letter} we study the cosmological constraints by
recent GRBs sample including 52 GRBs out to z$=$6.3 (Schaefer 2006).
As we mainly demonstrate the cosmological roles of GRBs, we focus on
$\Lambda$CDM models. There are three relevant parameters $H_0$,
$\Omega_m$ and $\OmegaL$. We concentrate on the constraints on (\Om
,\Ol). For $H_0$, instead of using a fixed value as in previous
studies, we treat it as a free parameter with a uniform prior in the
range of $H_0=72\pm8\enspace\kmsmpc$ to get more realistic
constraints from GRBs. We then combine GRBs with SNIa "gold" (Riess
et al. 2004)/SNLS (Astier et al. 2006) samples, the information from
WMAP three-year results (Spergel et al. 2006), and BAO from SDSS
(Eisenstein et al. 2005) to study their joint constraints.

\section{Observational Data}

We take the advantage of the recent GRB sample compiled by Schaefer
(2006) including 52 bursts with properly estimated and corrected
redshifts to investigate the cosmological constraints. The redshift
of the sample extends to z$=$6.3 with 25 objects having z$>$1.5.
Mosquera Cuesta et al. (2006) showed the distance modulus-redshift
relation of this sample explicitly. Upon using these distance
modulus, we fully aware of the circulation problem associated with
GRBs as cosmological probes (e.g. Firmani et al. 2006c). As the
correlations are calibrated in a cosmologically model-dependent way
due to the lack of a suitable set of low redshift GRBs, the
consistent analyses of the data should regard the cosmological
parameters as free ones in finding quantitative correlation
relations, and then study their constraints coherently (Ghirlanda et
al. 2006). On the other hand, it has been shown that the correlation
parameters (e.g. Firmani et al. 2006c) are not very sensitive to
cosmological models although the scatters do. Therefore we do not
expect large biases if using a given set of distance modulus
calibrated with a fiducial cosmological model to analyze their
cosmological constraints. In this paper, we take the distance
modulus calculated by Schaefer (2006; Mosquera Cuesta et al. 2006)
without any corrections for the circulation problem.

We also use SNIa gold sample and SNLS data in our analyses. The
Riess gold sample contains 157 data including 14 high redshift SNIa
with z$>$0.9 (Riess et al. 2004). The SNLS sample consists of 44
nearby (0.015$<$z$<$0.125) objects assembled from the literature,
and 73 distant SNIa (0.15$<$z$<$1.00) discovered and carefully
followed during the first year of SNLS (Astier et al. 2006). For the
cosmological fits, two of the SNLS data points were excluded because
they are outliers in the Hubble diagram. To analyze the role of GRB
and to study the combined constraints, we further make use of the
shift parameter $ {\cal R}=\sqrt{\Omega_m^0}\int_0^{z_{\rm
dec}}{dz\over E(z)}~,\label{R}$ determined from WMAP, CBI and ACBAR
(Wang \& Mukherjee 2006), and the parameter $A$ measured from SDSS
(Eisenstein et al. 2005) that is defined as
$A=\sqrt{\Omega_m^0}E(z_1)^{-1/3}\left[{1\over
z_1}\int_0^{z_1}{dz\over E(z)}\right]^{2/3}~\label{A}$ with
$z_1=0.35$. The CMB shift parameter contains the main information
for the scale of the first acoustic peak in the TT spectrum, and is
the most relevant one for constraining dark energy properties.
Recently, the new SDSS LRG data were released (Tegmark et al. 2006),
and the corresponding power spectrum was analyzed. The BAO peaks are
clearly seen in the power spectrum, which, together with the overall
shape, put tight constraints on model parameters.

For the fitting methodology, we use the standard \chs\ minimization
method. We also choose standard parametrization of $\Lambda$CDM
model instead of the non-parametric method (e.g. Daly \& Djorgovski
2004). It is known that parameter estimates depend sensitively on
the assumed priors on other parameters. In our study, we choose the
allowed range of the Hubble constant $H_0=72\pm8\enspace\kmsmpc$
resulting from the Hubble Space Telescope Key Project with a uniform
prior (Freedman et al. 2001), and marginalize over $H_0$ to get
two-dimensional constraints for \Om ~and \Ol . It is noted that most
of the previous work using GRBs to constrain cosmological parameters
took a fixed value of $H_0$.

\section{Complementarity of GRBs to Other Probes}

In this section, we study the cosmological constraints using data
sets we have discussed above.

Fig.1 shows confidence-level contours for the $\Lambda$CDM model
using the GRBs sample. We divide the sample into two sub-samples
with z$<$1.5 including 27 objects and z$>$1.5 including 25 objects,
respectively. The dot-dashed lines are the results from the z$>$1.5
sub-sample with the minimum $\chi^2=28.22$ at $\Omega_m=0.32$ and
$\OmegaL=0.93$. The dashed lines correspond to the z$<$1.5
sub-sample with the minimum $\chi^2=36.68$ at $\Omega_m=0.56$ and
$\OmegaL=1.48$. The solid lines show the constraints resulting from
the full GRBs sample with the minimum $\chi^2=65.78$ occurring at
\Om=0.43, and \Ol=0.91. We note that the degeneracy direction from
the z$<$1.5 sub-sample is similar to that from SNLS SNIa sample, but
the contours are much larger due to larger error bars in GRB data.
Thus with similar redshift distributions, GRBs cannot provide much
additional information in the parameter constraints in comparison
with that of SNIa. The constraints from the high-redshift sub-sample
have degeneracies rotating toward \Ol. We also show the results from
the full GRBs sample with a fixed Hubble constant
$H_0=74\enspace\kmsmpc$ (short dashed lines). The corresponding
minimum $\chi^2$ is $\chi^2=65.79$ at $\Omega_m=0.43$ and
$\OmegaL=0.91$. The contours are much more narrow than the solid
ones. This shows clearly that a strong prior can result an
overestimate on the power of a cosmological probe. It is also noted
that factitious priors on $H_0$ can result in strongly biased
constraints. For $H_0=68\enspace\kmsmpc$, the corresponding minimum
is $\chi^2=68.08$ at $\Omega_m=0.53$ and $\OmegaL=0.38$. For
$H_0=80\enspace\kmsmpc$, the corresponding minimum is $\chi^2=68.53$
at $\Omega_m=0.32$ and $\OmegaL=1.13$.

In Fig.2, we show the results from combining SNLS SNIa data with the
full GRBs sample and with the z$>$1.5 sub-sample, respectively. The
short dashed lines correspond to the constraints from the SNLS
sample. The dashed lines and the solid lines correspond to
constraints by combining SNLS with the z$>$1.5 GRB sub-sample (the
minimum $\chi^2=140.35$ at $\Omega_m=0.45$ and $\OmegaL=1.01$) and
with full GRBs sample (the minimum $\chi^2=181.62$ at
$\Omega_m=0.46$ and $\OmegaL=1.02$), respectively. It is seen
clearly that GRBs contribute considerably to the cosmological
constraints comparing with that from SNIa alone due to their much
more extended redshift range.

In Fig.3, we plot the results of the combined analysis of SNIa + BAO
with solid contours and compare it with the constraints from SNIa +
full GRB sample (dashed lines). We find that including the
information from BAO that tightly constrains the matter content \Om,
the contribution from GRBs is degraded. On the other hand, the two
sets of constraints are largely consistent with each other,
indicating the feasibility of using GRBs as cosmological probes.
Considering the central values of \Om~ and \Ol~, however, there
exists some differences between the results of SNIa+BAO and
SNIa+GRBs. These might imply the existence of some unknown
systematics for GRBs themselves as cosmological standard candles
and/or new phenomena for high-redshift universe. These apparent
differences deserve further investigations.


Fig.4 shows the confidence contours of the combined analyses
including CMB. The results from BAO alone, and from the CMB shift
parameter are shown in dotted and short dashed lines, respectively.
The constraints from the combined GRB+CMB+BAO and CMB+BAO+SNIa are
plotted, respectively in dot-dashed and dashed lines. The solid
lines are the constraints from CMB+BAO+SNIa+GRBs. The constraints
from GRB+CMB+BAO are less restrictive than, but consistent with
those from CMB+BAO+SNIa. The concordance model is well within the
68.4\% confidence level.

\section{Discussion and Conclusion}

In this \emph{letter}, we study the role of GRBs in constraining
cosmological parameters using the recent GRB sample including 52
GRBs out to z$=$6.3 (Schaefer 2006). We study the constraints using
GRBs alone, and further focus on their complementarity to other
cosmological probes. In our analyses, We adopt $H_0$ as a free
parameter in the range of $H_0=72\pm8\enspace\kmsmpc$. In comparison
with the results with a strong prior on $H_0$ (a fixed value of
$H_0$)(Firmani et al. 2006c), the constraints by GRBs alone are much
less restrictive. However, the complementarity of GRBs to SNIa are
clearly seen, which is mainly contributed by the sub-sample of GRBs
with z$>$1.5. This demonstrates the usefulness of high redshift
cosmological probes. On the other hand, as BAO highly constraints
\Om, the contribution of GRBs is degraded by including BAO
information in the analyses. The combined analyses of
SNIa+GRB+CMB+BAO result constraints on \Om~and \Ol that are broadly
consistent with the concordance $\Lambda$CDM model.

We notice the differences between the central values of (\Om, \Ol)
from GRBs+SNIa and those from SNIa+BAO. At this stage, we are not
able to draw strong conclusions on these because of the large
confidence contours. If the differences persist as each of the
constraints improves, they might indicate some unknown systematics
involved in GRBs, or our knowledge about our universe might not be
complete.

The feasibility of GRBs as cosmological probes bridges up the gap
between the relatively nearby SNIa and CMB. At present, the
quantitative correlations discovered for GRBs are cosmologically
model dependent, and therefore there is a notorious circulation
problem in applying these correlations to constrain cosmologies. It
is thus crucial to find large enough number of low redshift GRBs (or
GRBs in a narrow redshift range) to calibrate the correlations in a
model-independent way. On the other hand, to understand the GRB
physics more thoroughly may lead to definite correlations between
different observables without involving cosmological models. The
information on the redshift distribution of GRBs is also of great
importance in cosmological studies. Recently, Le and Dermer (2006)
found that with the assumption of continued positive evolution of
the GRBs rate density to high redshift, the fraction of
high-redshift GRBs is estimated to be 8-12\% and 2.5-6\% at z $\geq$
5 and z $\geq$ 7, respectively. With the cumulation of observational
data with high qualities, we expect great advances in all this
important aspects, and therefore GRBs can provide us much more
cosmological information than we have today.

Flat geometry with a relic cosmological term ($\Lambda$CDM) seems to
be in agreement with almost all the cosmological observations. From
the theoretical viewpoint, however, it faces the well-known headache
cosmological constant problem (e.g. Weinberg 1989). This and other
difficulties motivate different categories of dynamical dark energy
models. Current observations provide us only limited knowledge on
dark energy, especially on its evolutionary properties. It is
therefore of great interests to explore high redshift cosmological
probes. GRBs are the only possible candidates for the purpose. In
this letter, we concentrate on demonstrating the cosmological roles
of GRBs and only present the constraints on \Om~and \Ol~ in
$\Lambda$CDM cosmologies. In our forthcoming paper, we will show the
results of global fitting and emphasize the contribution of GRBs to
the constraints on the dynamics of dark energy. In short, with
limited information from medium and low redshift standard candles
such as SNIa so far, high-redshift information from GRBs can help us
to shrink the allowed cosmological parameter space, and might
further find some interesting behaviors of dark energy at high
redshifts. On the other hand, it is theoretically desirable to find
out a suitable parametrization of dark energy that could reflect its
high-redshift behaviors properly and effectively.

SNIa has been proved playing a central role in elucidating the
nature of dark energy. High redshift GRBs as extensions of such
standard-candle-like probes are promising and important for
cosmological studies. Similar to the anchor for the Hubble diagram
provided by low redshift SNIa with z$\sim$0.05 (Linder 2006), GRBs
could perform as a high-redshift hoop helping us to calibrate the
middle part of the Hubble diagram more precisely, and thus to get
better constraints on the nature of dark energy. Furthermore, high
redshift GRBs could directly test exotic dark energy models such as
the oscillating dark energy model (Linder 2005) or bump like models
(Xia et al. 2004). With CMB from redshift around 1100, 21cm emission
and GRBs covering high redshift up to around 80, and SNIa, BAO, weak
lensing and Integrated Sachs Wolfe (ISW) effect providing medial and
low redshift information, the global picture of the cosmic evolution
could be obtained.

\acknowledgments{M.S. acknowledges valuable discussions with Dai
Zigao, Li Hong, Chen Xuelei, Yue Youling and technical supports from
Wang Zheng. This research was supported in part by the National
Science Foundation of China under grants 10243006 and 10373001, by
the Ministry of Science and Technology of China under grant
TG1999075401, by the Key Grant Project of Chinese Ministry of
Education (305001), and by the National Science Foundation of China
under grant 10533010.}

\clearpage
\newpage

\newpage
\clearpage

\begin{figure}
\begin{center}
\resizebox{17cm}{13cm}
{\includegraphics{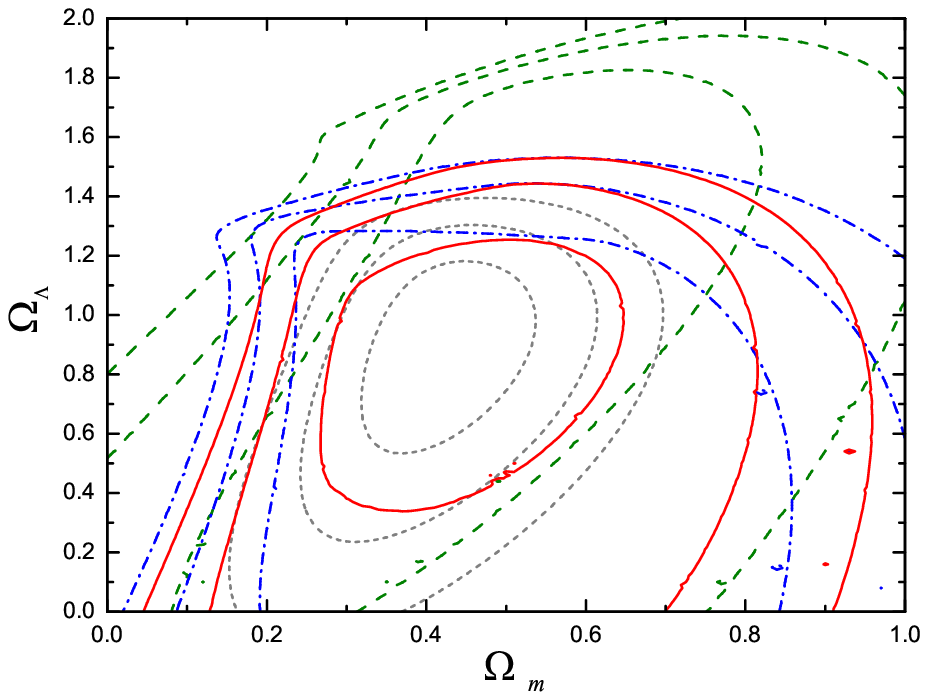}} \figcaption[]{1, 2, and 3 $\sigma$
confidence level contours on ($\Omega_m$, $\OmegaL$) using the GRB
sample. We adopt $H_0=72\pm8\enspace\kmsmpc$. The dot-dashed lines
are the results from the z$>$1.5 sub-sample with the minimum
$\chi^2=28.22$ at $\Omega_m=0.32$ and $\OmegaL=0.93$. The dashed
lines correspond to the z$<$1.5 sub-sample with the minimum
$\chi^2=36.68$ at $\Omega_m=0.56$ and $\OmegaL=1.48$. The solid
lines show the constraints resulting from the full GRBs sample with
the minimum $\chi^2=65.78$ occurring at \Om=0.43, and \Ol=0.91. The
short dashed lines correspond to the full sample constraints with
fixed $H_0=74\enspace\kmsmpc$. The corresponding minimum $\chi^2$ is
$\chi^2=65.79$ at $\Omega_m=0.43$ and $\OmegaL=0.91$. \label{fig1}}
\end{center}
\end{figure}

\clearpage
\newpage
\clearpage

\begin{figure}
\begin{center}
 \resizebox{17cm}{13cm}
{\includegraphics{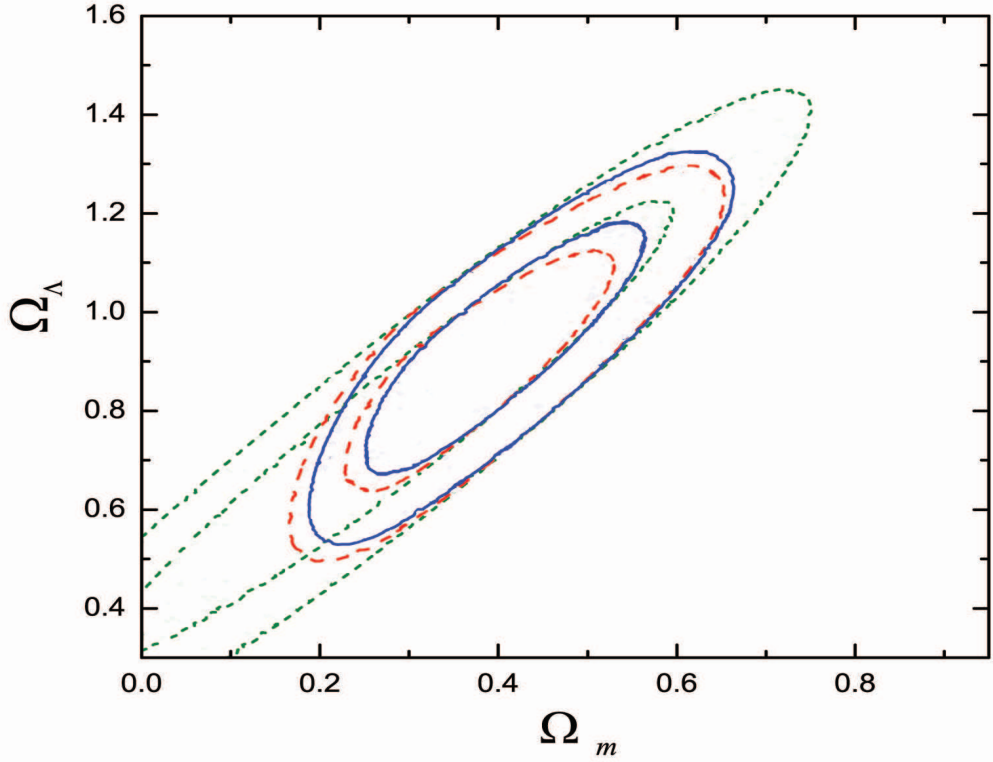}} \figcaption[]{{1, 2, and 3 $\sigma$
confidence level contours for ($\Omega_m$, $\OmegaL$). The short
dashed lines correspond to SNLS sample. The dashed lines and the
solid lines correspond to constraints combining SNLS with the
z$>$1.5 GRB sub-sample with minimum $\chi^2=140.35$ at
$\Omega_m=0.45$ and $\OmegaL=1.01$, and with full GRBs sample with
minimum $\chi^2=181.62$ at $\Omega_m=0.46$ and $\OmegaL=1.02$,
respectively.} \label{fig2}}
\end{center}
\end{figure}

\newpage
\clearpage

\begin{figure}
\begin{center}
\resizebox{17cm}{13cm}
{\includegraphics{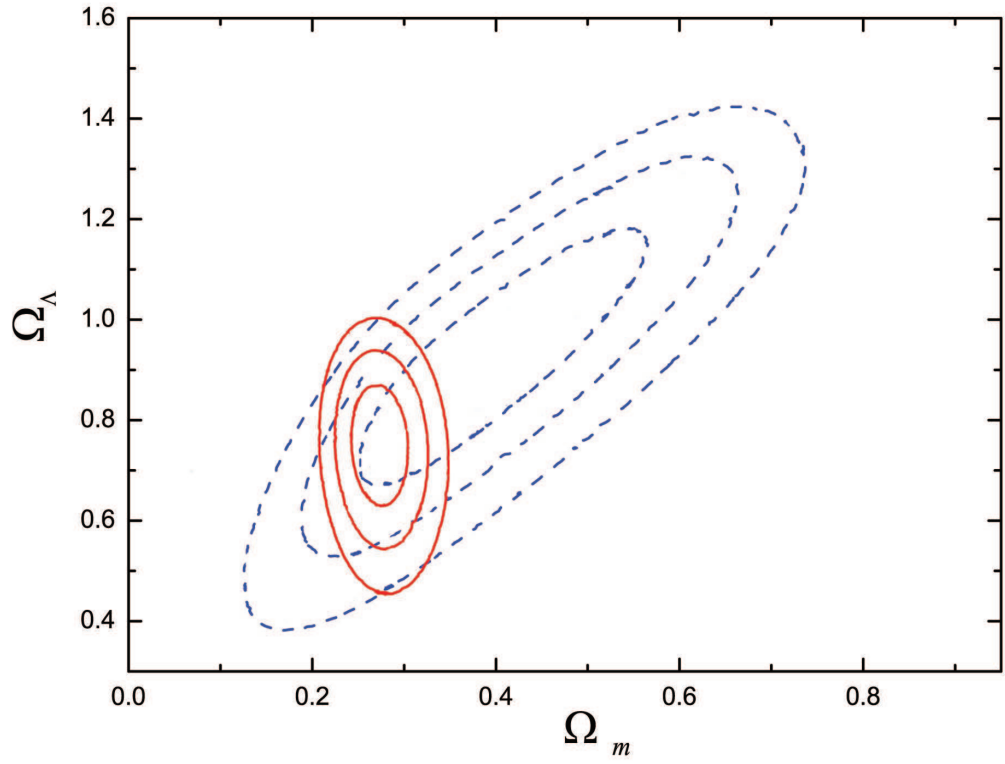}} \figcaption[]{1, 2, and 3 $\sigma$
confidence level contours for ($\Omega_m$, $\OmegaL$). The solid
contours show combined analysis of SNIa + BAO, and the constraints
from SNIa + full GRBs sample are shown in dashed lines.\label{fig3}}
\end{center}
\end{figure}


\clearpage

\newpage
\clearpage

\begin{figure}
\begin{center}
 \resizebox{17cm}{13cm}
{\includegraphics{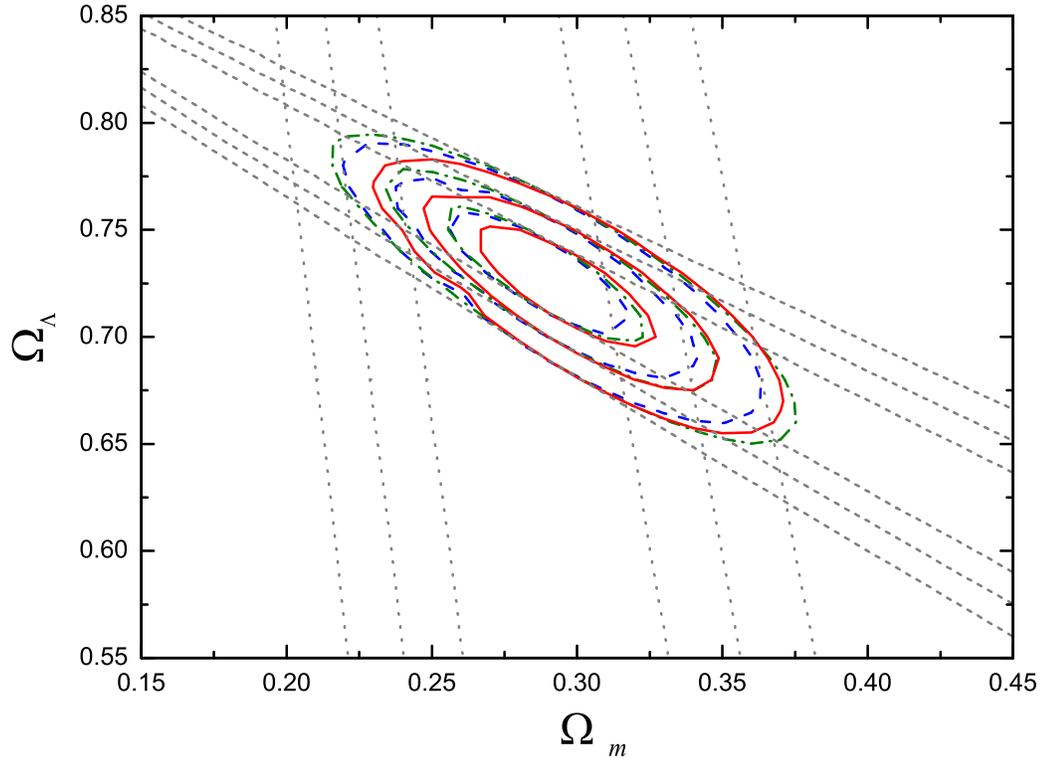}} \figcaption[]{{1, 2, and 3 $\sigma$
confidence level contours for ($\Omega_m$, $\OmegaL$). The dotted
lines show constraints from BAO alone, short dashed lines lay out
constraints by the CMB shift parameter alone. The combined
GRB+CMB+BAO constraints are shown with dot-dashed contours. The
dashed contours correspond to CMB+BAO+SNIa and the solid contours
correspond to the combined constraints from CMB+BAO+SNIa+GRBs.}
 \label{fig4}}
\end{center}
\end{figure}


\end{document}